# Stabilizing effect of enhanced resistivity on peeling-ballooning instabilities on EAST

X Lin[1,2], D Banerjee[2], P. Zhu[3,4,2*], G. S. Xu[1*], Y. Ye[5], Y. F. Wang[1], Q. Zang[1], T. Zhang[1], and Y. J. Chen[1]

[1] Institute of Plasma Physics, Chinese Academy of Sciences, Hefei, Anhui 230031, China

[2] University of Science and Technology of China, Hefei, Anhui 230026, China

[3] International Joint Research Laboratory of Magnetic Confinement Fusion and Plasma Physics, State Key Laboratory of Advanced Electromagnetic Engineering and Technology, School of Electrical and Electronic Engineering, Huazhong University of Science and Technology, Wuhan, Hubei 430074, China

[4] Department of Engineering Physics, University of Wisconsin-Madison, Madison, Wisconsin 53706, USA

[5] Shenzhen University, Shenzhen, Guangdong 518060, China

Corresponding author: zhup@hust.edu.cn and gsxu@ipp.ac.cn

**Abstract**

Previous stability analysis of NSTX equilibrium with lithium-conditioning demonstrates that the enhanced resistivity due to the increased effective charge number $Z_{eff}$ (i.e. increased impurity level) can provide a stabilizing effect on low-n edge localized modes (Banerjee *et al* 2017 *Nucl. Fusion* **24** 054501). This paper extends the resistivity stabilizing effect to the intermediate-n peeling-ballooning (PB) instabilities with the linear stability analysis of EAST high-confinement mode equilibria in NIMROD two-fluid calculations. However, the resistivity stabilizing effect on PB instabilities in the EAST tokamak appears weaker than that found in NSTX. This work may give better insight into the physical mechanism behind the beneficial effects of impurity on the pedestal stability.

## 1. Introduction

Edge localized mode (ELM) is a critical issue in magnetic confinement fusion research, especially for future fusion reactors, such as the International Thermonuclear Experimental Reactor (ITER) [1] and the China Fusion Engineering Test Reactor (CFETR) [2]. According to the peeling-ballooning (PB) theory [3, 4], ELMs are expected to be triggered by the coupled peeling-ballooning modes (PBMs) which can lead to the periodic bursts of heat and particle fluxes onto divertor. These transient fluxes can result in a long-term erosion of plasma-facing components (PFCs), especially the divertor target plates [5]. Furthermore, the sputtering of plasma-facing wall materials can cause impurity accumulation in the core plasma and deteriorate the energy confinement [6, 7]. In recent years, various technologies aiming to mitigate the heat loads of large ELMs have been developed, such as impurity injection [8], supersonic molecular beam injection [9], resonant magnetic perturbation [10], etc. The external impurity injection is one of the highly promising methods for ELM mitigation or suppression demonstrated on multiple fusion devices [8, 11-13]. In addition, the enhanced pedestal pressure and improved energy confinement were also observed with impurity injection [11, 12, 14, 15].

Previous efforts have been made on the understanding of the physical mechanism of impurity effect both in theory and in experiments. Earlier studies have focused on the physical effects of profile modification due to impurity injection on the pedestal magnetohydrodynamic (MHD) stability [11, 12, 16, 17]. It is worth noting that the phenomenon of increased effective charge number $Z_{eff}$, which represents a higher impurity level in plasma, has been widely observed in the impurity injection experiments in many devices, such as ASDEX Upgrade [15], JET [14], DIII-D [11], NSTX [18] and JT-60U [19]. One of the impurity effects is the reduction of main ion density by dilution, which is guessed to play a role in improving PB stability and confinement [11, 14, 15]. Besides the dilution effect, a recent linear stability analysis of NSTX equilibria with the initial-value NIMROD code reveals that the enhanced resistivity affected by $Z_{eff}$ could also provide an additional stabilizing effect on low-n edge localized modes [20, 21].

The NIMROD calculations of NSTX equilibria give a new insight into the physical mechanisms of impurity effects. However, the existence of this resistive stabilization effect in EAST had been unknown. In this paper, to further study the resistive stabilization effect in EAST, linear stability analysis of two ELMy equilibria is carried out with the two-fluid extended MHD model in NIMROD code, which has been benchmarked and verified for the simulations of both ideal and non-ideal physical processes [22-30]. Our calculations find that the enhanced resistivity due to increased $Z_{eff}$ can provide a stabilizing effect on both low-n and intermediate-n PBMs, however this stabilizing effect in EAST appears weaker than that observed in NSTX.

The rest of the paper is organized as follows. In section 2, the two-fluid extended MHD equations in NIMROD and the EAST ELMy equilibria are described. Then the main results of stabilizing effect of increasing $Z_{eff}$ on PBMs are reported in section 3. Finally, a brief conclusion is made in section 4.

**2. Extended MHD model and equilibria**

To simulate the realistic plasma and assess the impact of $Z_{eff}$ on the PB instabilities, the two-fluid extended MHD model in NIMROD [31], including the non-ideal effects such as resistivity, finite-Larmour-radius (FLR) effect, two-fluid Hall and electron diamagnetic drift, is adopted in our linear simulation. The two-fluid extended MHD equations used in our NIMROD calculations are

$$\frac{\partial n}{\partial t} + \nabla \cdot (n\mathbf{u}) = 0, \quad (1)$$

$$mn\left(\frac{\partial}{\partial t} + \mathbf{u}\cdot\nabla\right)\mathbf{u} = \mathbf{J}\times\mathbf{B} - \nabla p - \nabla\cdot\overline{\Pi}, \quad (2)$$

$$\frac{3}{2}n\left(\frac{\partial}{\partial t} + \mathbf{u}_\alpha\cdot\nabla\right)T_\alpha = -nT_\alpha\nabla\cdot\mathbf{u}_\alpha - \nabla\cdot\mathbf{q}_\alpha \quad (\alpha = i, e), \quad (3)$$

$$\frac{\partial \mathbf{B}}{\partial t} = -\nabla\times\left[\eta\mathbf{J} - \mathbf{u}\times\mathbf{B} + \frac{1}{ne}(\mathbf{J}\times\mathbf{B} - \nabla p_e)\right], \quad (4)$$

$$\mu_0 \mathbf{J} = \nabla\times\mathbf{B}, \quad \nabla\cdot\mathbf{B} = 0, \quad (5)$$

Where $n$ is number density, $\mathbf{u}$ is the center-of-mass flow velocity, $m$ is ion mass, $\mathbf{J}$ and $\mathbf{B}$ are current density and magnetic field respectively, $p$ denotes the total plasma pressure including ion pressure $p_i$ and electron pressure $p_e$, $\mathbf{q}_{i,e}$ are conductive heat flux vectors of ion and electron, $\eta$ is the resistivity, and $\overline{\Pi}$ is the ion stress tensor including gyro-viscous components which have been described in [22].

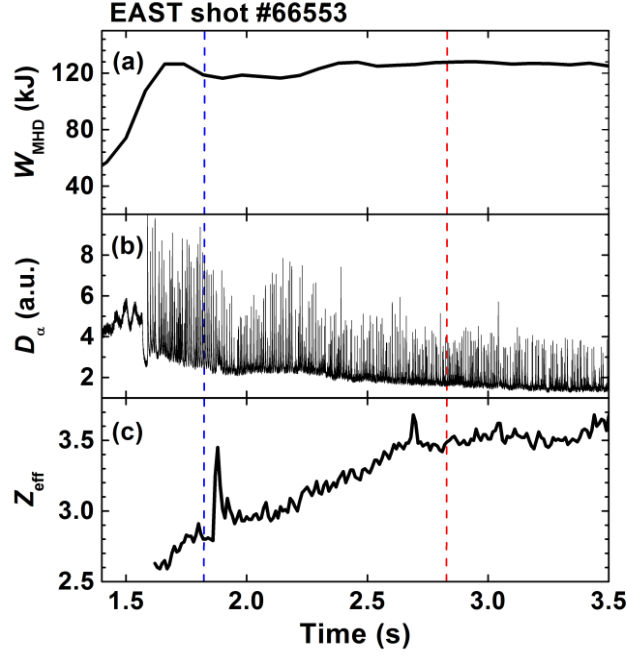

**Figure 1.** Time traces of (a) stored energy, (b) $D_\alpha$ emission from divertor region and (c) central line-averaged effective ion charge $Z_{eff}$ in the EAST discharge #66553.

In order to study the resistivity stabilizing effect in EAST, the high-confinement mode (H-mode) discharge #66553 with plasma current $I_p$ = 0.45 MA, toroidal field $B_t$ = 2.5 T, $\beta_p$ = 1, edge safety factor $q_{95}$ = 6.3, triangularity $\delta$ = 0.5 and elongation $\kappa$ = 1.60, as shown in figure 1, is chosen in our simulation. No external impurity is injected in the discharge. However, after L-H transition, the core line-averaged effective ion charge $Z_{eff}$ measured by visible bremsstrahlung system [32] increases significantly from 2.6 to 3.5. The increasing $Z_{eff}$ can be associated with the intrinsic impurity in the plasma. In this work, we carried out the linear stability analysis of equilibrium 1 with relatively lower $Z_{eff}$ ~ 2.8 at about 1880 ms, which is operated with line-averaged density $<n_e>$ ~ $3.5\times10^{19}$ m$^{-3}$, 2.45 GHz lower hybrid current drive (LHCD) power $P_{2.45G,LHCD}$ = 0.5 MW, 4.6 GHz LHCD power $P_{4.6G,LHCD}$ = 2 MW, electron cyclotron resonance heating (ECRH) power $P_{ECRH}$ = 0.3 MW, and equilibrium 2 with higher $Z_{eff}$ ~ 3.5 at about 2880 ms, which is operated with $<n_e>$ ~ $3.3\times10^{19}$ m$^{-3}$, same LHCD and ECRH powers and added 1.4 MW ion cyclotron resonance heating (ICRH) power. From the $D_\alpha$ emission in figure 1(b), the ELM at about 2880 ms appears to be smaller than that at 1880 ms. More detailed description of this experiment is given in [33].

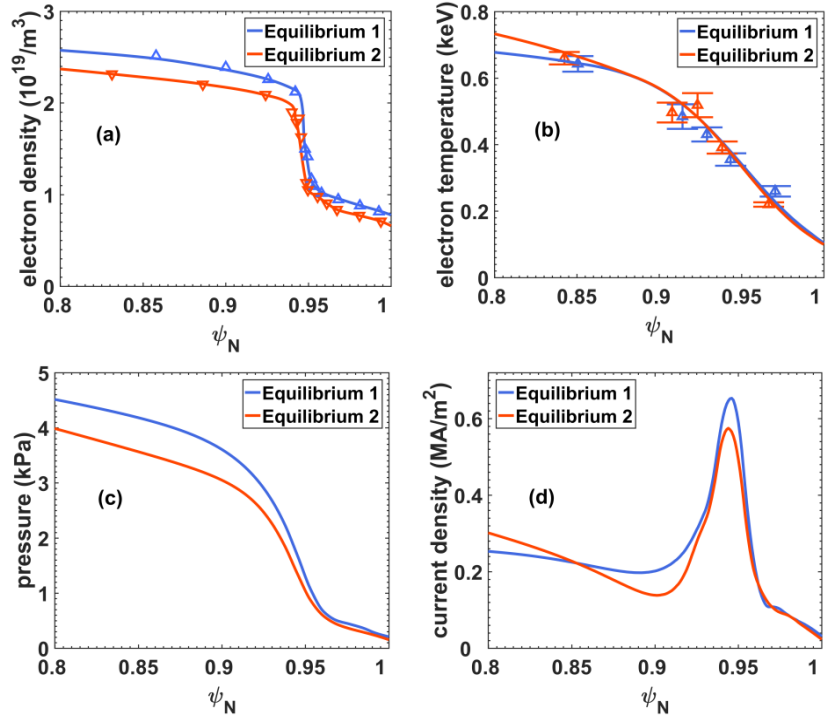

**Figure 2.** Radial profiles of (a) electron density, (b) electron temperature, (c) total pressure and (d) current density of the EAST H-mode discharge #66553 for equilibrium 1 (blue lines) and equilibrium 2 (red lines). $\psi_N$ is normalized poloidal flux.

Figure 2 shows the electron density, electron temperature, total plasma pressure and current density profiles of the two equilibria. The edge electron density profiles are measured by microwave reflectometer system [34] which is located at low-field side midplane. The electron temperature profiles are obtained from Thomson Scattering system [35, 36]. The total pressure profiles are the sum of electron pressure, ion pressure, and fast ion pressure. The edge flux-surface-averaged bootstrap current density profiles are calculated using Sauter model [37, 38]. In the comparison of the two equilibria, the pedestal electron temperatures are much similar. However, the pedestal density of equilibrium 1 is significantly higher than that of equilibrium 2, which directly leads to a higher pedestal pressure. In addition, the equilibrium 1 has a higher bootstrap current density probably due to its slightly steeper pedestal pressure and lower $Z_{eff}$. With the assumption that the edge ion temperature is equal to electron temperature, the two equilibria are generated with EFIT code [39] within the constraints of these experimental profiles.

## 3. Calculation results

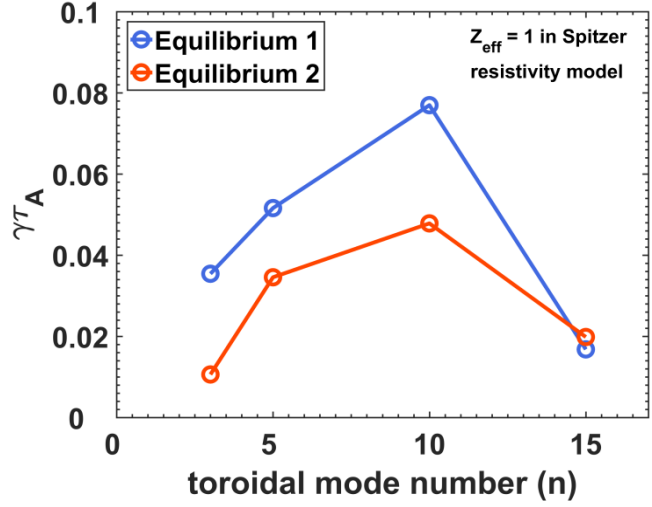

**Figure 3.** Normalized growth rates computed by NIMROD in two-fluid MHD model with equilibrium 1 (blue) and equilibrium 2 (red).

The linear stability analysis of PB instability was conducted in NIMROD two-fluid MHD model, as shown in figure 3. It is seen that the $n$ = 3-15 modes of the two cases are all unstable, and their normalized growth rates are both peaked at $n$ = 10. Nevertheless, the equilibrium 1 is more unstable than the other one, which is likely due to the steeper pedestal pressure gradient and higher edge current density. In addition, since FLR effect has a significant stabilization effect on high-n modes [29, 40], the growth rate of $n$ = 15 mode is strongly reduced relative to $n$ = 5 and 10 modes in the two cases. In this calculation, we utilized the realistic Spitzer resistivity model defined as

$$\eta\left(T_e\right) = Z_{eff}\eta_0 \left(\frac{T_{e0}}{T_e}\right)^{3/2}, \quad (6)$$

where $Z_{eff}\eta_0$ and $T_{e0}$ denote the resistivity and electron temperature at the magnetic axis respectively. In this model, the Spitzer resistivity is a function of electron temperature so that the resistivity is very low in the core plasma and increases dramatically in the pedestal region due to the drop of electron temperature. $Z_{eff}$ is also an important parameter that affects the edge resistivity in the model. However, the $Z_{eff}$ profile is not available due to diagnostic constraints on EAST. In this work, the $Z_{eff}$ is assumed to be uniform in the whole plasma. For the results in figure 3, the influence of resistivity contributed from $Z_{eff}$ on the PB instabilities is eliminated by setting $Z_{eff}$ equal to 1 in the Spitzer resistivity model. In the following, we scan the resistivity through changing $Z_{eff}$ in the Spitzer resistivity model for the two equilibria to study the impact of resistivity on the PB instabilities.

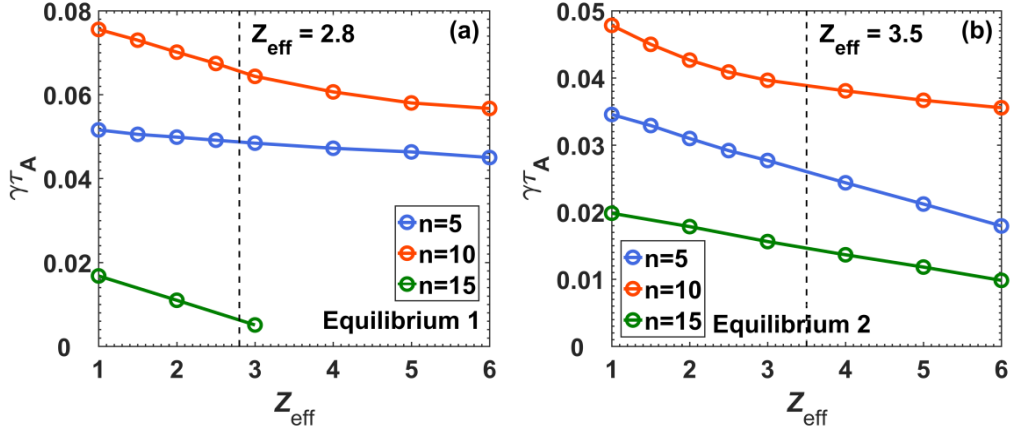

**Figure 4.** Normalized growth rates of $n$ = 5, 10, 15 modes as functions of $Z_{eff}$ computed by NIMROD two-fluid model with (a) equilibrium 1 and (b) equilibrium 2.

Based on the Spitzer resistivity model and the two-fluid MHD equations, a scan of $Z_{eff}$, i.e. resistivity, demonstrates that the growth rates of $n$ = 5, 10 and 15 modes decrease with increasing resistivity for the two equilibria, as shown in figure 4. This result confirms that the increased $Z_{eff}$ can also provide a stabilizing effect on both low-n and intermediate-n PB instabilities, which may be beneficial for stabilization of ELMs. This stabilizing effect appears stronger overall in equilibrium 2 with higher experimental $Z_{eff}$, therefore, it should be reasonable to expect that this stabilizing effect increases as the experimental $Z_{eff}$ rising from a low value to a higher one. Overall speaking, the stabilizing effect of increasing resistivity could be comparable to the stabilization of PBMs from profile modifications between the two equilibria, which suggests that the variation of pedestal profile and the enhanced $Z_{eff}$ could be two major factors affecting the PB instabilities.

It is important to note that the stabilizing effect of enhanced resistivity on NSTX is evident only for the low-n modes. In contrast, the calculations in the present work extend the resistive stabilization effect to the intermediate-n modes. The above results verify the universality of the resistive stabilization effect on PBMs, however, this stabilizing effect of enhanced resistivity in EAST appears weaker than that observed in NSTX, where the low-n modes are fully stabilized with increasing $Z_{eff}$ even for an ELMy case [20, 21].

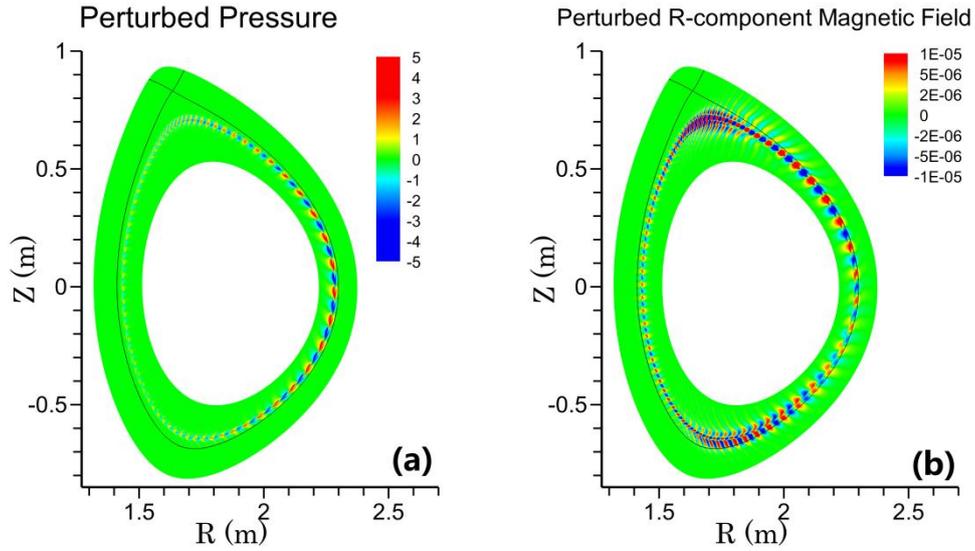

**Figure 5.** Contour plots of (a) perturbed pressure and (b) perturbed R-component magnetic field of the most unstable mode ($n = 10$) for equilibrium 2.

Figure 5 shows the contour plots of perturbed pressure and perturbed R-component magnetic field of the most unstable mode ($n = 10$) for equilibrium 2. The spatial structure of the PBM is well resolved numerically. As shown in figure 6, good convergences for four key numerical parameters have been demonstrated. To eliminate the errors from inconsistent parameter values, the radial and poloidal grid point numbers, the time step and the polynomial degree remain same for all the above calculations shown in figures 3-5, which are set to be 72, 468, $1 \times 10^{-8}$, and 5 respectively.

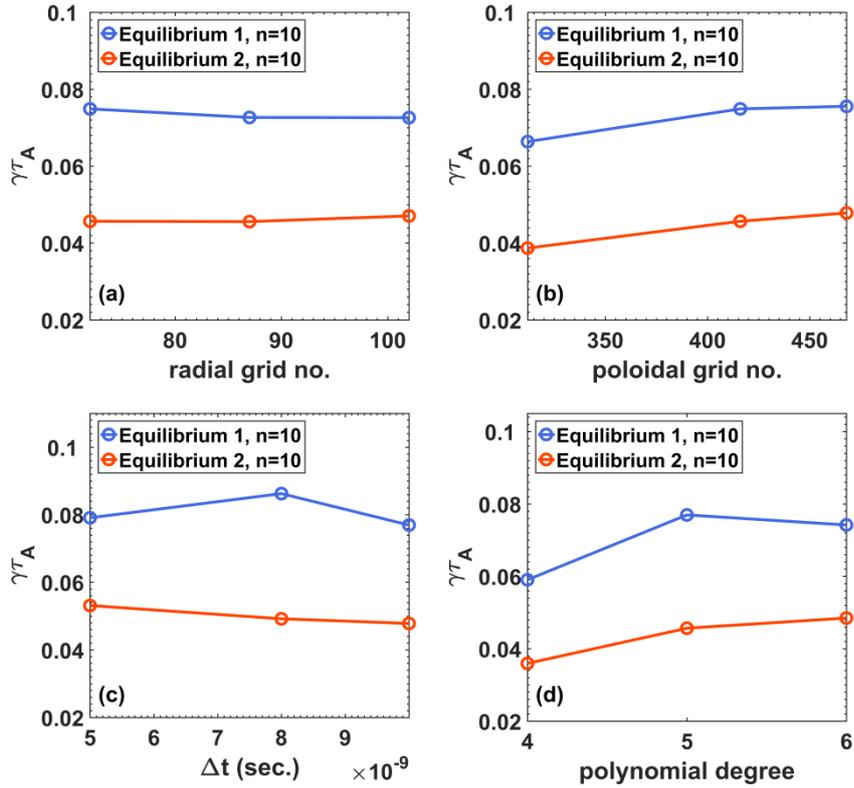

**Figure 6.** Normalized growth rates of n = 10 mode as functions of (a) radial and (b) poloidal grid points, (c) time step and (d) polynomial degree of finite elements used in NIMROD code.

## 4. Conclusions

In this paper, linear stability analysis of two EAST H-mode equilibria with different $Z_{eff}$ is carried out by solving the extended two-fluid MHD equations using NIMROD code. The calculations show that the most unstable modes of the two equilibria are both $n = 10$ mode and the high-n modes are strongly stabilized by FLR effect included in the two-fluid MHD model. For $Z_{eff} = 1$, equilibrium 2 is more stable than the other one due to the variations of edge profiles. Based on the Spitzer resistivity model, the scanning of resistivity through increasing $Z_{eff}$ indicates that the enhanced resistivity can provide a stabilizing effect on both low-n and intermediate-n peeling-ballooning instabilities, which could be a candidate for explaining the beneficial effect of impurity on the pedestal stability in EAST. Nevertheless, this resistivity stabilizing effect on PB instabilities in EAST appears weaker than that observed in NSTX.

The resistivity effect is closely correlated with the impurity level in the edge region. Particularly, in the impurity injection experiments, substantial low-Z impurities will accumulate in the plasma edge region, which leads to significant increases in edge $Z_{eff}$ and resistivity and thus contributes to the stabilization of PB instabilities. Therefore, the resistivity stabilizing effect on the PB instabilities should be non-negligible in the analysis of impurity injection experiments. Future stability analysis should also take into account of non-uniform radial distributions of impurity species and $Z_{eff}$.

**Acknowledgement**

This work was supported by National Magnetic Confinement Fusion Science Program of China under Contracts Nos. 2014GB124002, 2015GB101000, 2015GB101004, National Natural Science Foundation of China under Contracts Nos. 11575235, 11575236, 11705237, 11775221, the Fundamental Research Funds for the Central Universities at Huazhong University of Science and Technology Grant No. 2019kfyXJJS193, U.S. Department of Energy Grant Nos. DE-FG02-86ER53218 and DE-SC0018001. We are grateful for the support from the NIMROD team. This research used resources of the National Energy Research Scientific Computing Center, a DOE Office of Science User Facility supported by the Office of Science of the U.S. Department of Energy under Contract No. DE-AC02-05CH11231.